\documentclass[twocolumn,epjc3]{svjour3}
\smartqed
\usepackage{graphicx}
\usepackage{mathptmx}
\usepackage{amsmath}
\usepackage{microtype}
\usepackage[colorlinks,citecolor=blue,urlcolor=blue,linkcolor=blue]{hyperref}
\usepackage[numbers,sort&compress]{natbib}
\usepackage{orcidlink}
\usepackage{listings}
\usepackage{multirow}
\usepackage{xcolor}

\definecolor{codegreen}{rgb}{0,0.6,0}
\definecolor{codegray}{rgb}{0.5,0.5,0.5}
\definecolor{codepurple}{rgb}{0.58,0,0.82}
\definecolor{backcolour}{rgb}{0.95,0.95,0.92}

\lstdefinestyle{mystyle}{
    backgroundcolor=\color{backcolour},   
    commentstyle=\color{codegreen},
    keywordstyle=\color{magenta},
    numberstyle=\tiny\color{codegray},
    stringstyle=\color{codepurple},
    basicstyle=\ttfamily\footnotesize,
    breakatwhitespace=false,         
    breaklines=true,                 
    captionpos=b,                    
    keepspaces=true,                 
    numbers=left,                    
    numbersep=5pt,                  
    showspaces=false,                
    showstringspaces=false,
    showtabs=false,                  
    tabsize=2
}

\journalname{Eur. Phys. J. A}
\begin{document}

\title{\texttt{NuLattice}: \emph{Ab initio} computations of atomic nuclei on lattices\thanksref{t1}
}

\author{
M.~Rothman\orcidlink{0000-0001-9991-5670}\thanksref{addr1,e1}
\and
B.~{Johnson-Toth}\orcidlink{0009-0003-5525-0357}\thanksref{addr1,e2}
\and
G.~Hagen\orcidlink{0000-0001-6019-1687}\thanksref{addr3,addr1,e5}
\and
M.~Heinz\orcidlink{0000-0002-6363-0056}\thanksref{addr3,addr4,e3}
\and 
T.~Papenbrock\orcidlink{0000-0001-8733-2849}\thanksref{addr1,addr3,e4}
}

\thankstext{e1}{\email{mrothma1@vols.utk.edu}}
\thankstext{e2}{\email{bjohn214@vols.utk.edu}} 
\thankstext{e5}{\email{hageng@ornl.gov}}
\thankstext{e3}{\email{heinzmc@ornl.gov}}
\thankstext{e4}{\email{tpapenbr@utk.edu}}

\thankstext{t1}{This manuscript has been authored in part by UT-Battelle, LLC, under contract DE-AC05-00OR22725 with the US Department of Energy (DOE). The US government retains and the publisher, by accepting the article for publication, acknowledges that the US government retains a nonexclusive, paid-up, irrevocable, worldwide license to publish or reproduce the published form of this manuscript, or allow others to do so, for US government purposes. DOE will provide public access to these results of federally sponsored research in accordance with the DOE Public Access Plan (\url{http://energy.gov/downloads/doe-public-access-plan}).}

\institute{
\label{addr1}
Department of Physics and Astronomy, University of Tennessee, Knoxville, Tennessee 37996, USA 
\and
\label{addr3}
Physics Division, Oak Ridge National Laboratory, Oak Ridge, Tennessee 37831, USA 
\and
\label{addr4}
National Center for Computational Sciences, Oak Ridge National Laboratory, Oak Ridge, Tennessee 37831, USA 
}

\date{}

\maketitle

\begin{abstract}
We introduce \texttt{NuLattice}, a Python software package for \textit{ab initio} computations of atomic nuclei on lattices. The computational tools consist of Hartree Fock, the coupled cluster method, the in-medium similarity renormalization group, and full configuration interaction. At present, the employed interactions are from pion-less effective field theory at leading order and consist of two-body and three-body contacts. We present results for light nuclei $^{2}$H, $^{3,4}$He, $^{8}$Be, $^{12}$C, and $^{16}$O. \texttt{NuLattice} algorithms exploit the sparsity and locality of lattice interactions, and as a result computations can be run on laptops.      
\end{abstract}

\section*{Program Summary and Specifications}
\begin{description}[font=\normalfont\itshape]
  \item[Program title:] \texttt{NuLattice}
  \item[Licensing provisions:] BSD-3-clause
  \item[Programming language:] Python
  \item[Repository and DOI:] \href{https://github.com/NuLattice}{\nolinkurl{github.com/NuLattice}}\\ 
  DOI: \href{https://doi.org/10.5281/zenodo.17094172}{\nolinkurl{10.5281/zenodo.17094172}}
  \item[Description of problem:] \textit{Ab initio} computations of atomic nuclei require considerable computing resources and a significant effort to generate and process the nuclear interactions used as input.
  \item[Method of solution:] Using a lattice in position space significantly simplifies the generation of nuclear interactions and, due to the short range of the nuclear force, allows computing techniques to exploit sparse data structures for efficient implementations. This program  package contains full configuration interaction for light nuclei, Hartree Fock, coupled-cluster method, and the in-medium similarity renormalization group to solve the nuclear many-body problem. It can be used in research and education.
\end{description}

\sloppy
\section{Introduction}
\label{sec:intro}

\textit{Ab initio} computations of atomic nuclei, i.e., computations that use Hamiltonians from effective field theories of quantum chromodynamics and only make controlled approximations~\cite{Ekstrom:2022yea}, have made significant progress over the last decade~\cite{elhatisari2015,hagen2015,piarulli2018,morris2018,yao2020,arthuis2020,dytrych2020,maris2021,stroberg2021,hu2022,elhatisari2024,sun2025,Door2025,Bonaiti2025}. Reference-state based methods such as the no-core shell model~\cite{navratil2009,barrett2013}, coupled-cluster theory~\cite{coester1960,kuemmel1978,bishop1991,hagen2014}, the in-medium similarity renormalization group (IMSRG)~\cite{tsukiyama2011,hergert2016}, and Green's function approaches~\cite{dickhoff2004,soma2013} usually employ a harmonic oscillator basis. This is in contrast to nuclear lattice effective field theory~\cite{lee2009,lahde2019}, lattice  approaches to few-body systems~\cite{Konig:2020lzo,Konig:2022cya}, and mean-field computations~\cite{Bennaceur:2005mx,bonche2005,pei2014,Reinhard:2021yke,Jin:2020kdn,Chen:2021cpe}, which are based on spatial lattices. 

In this paper we use a spatial lattice and employ full configuration interaction, coupled-cluster theory, IMSRG, and Hartree Fock to solve nuclei based on Hamiltonians from pion-less effective field theory~\cite{bedaque2002,kirscher2010,lensky2016,konig2016}. We do so by using the newly developed Python package \texttt{NuLattice} that we also present in this work. The lattice approach offers one key advantage over the usual harmonic oscillator basis. It exposes the short range of nuclear forces and tremendously reduces the number of matrix elements that need to be stored and processed. As a consequence, one does not need tens of terabytes to store the matrix elements of the three-nucleon forces~\cite{miyagi2022}.  

This makes it then possible to use Python as a programming language and to perform nontrivial \textit{ab initio} computations with a laptop. 
This could make \textit{ab initio} computations accessible to a larger group of researchers and also to educators and students. Inspired by the open-source software~\cite{Stroberg_IMSRG_2018, Arthuis:2018yoo, psi4_2020,pyscf2020,bally2021b, Tichai:2021ewr, miyagi2023EPJA},  \texttt{NuLattice} is publicly available. 

\section{Using lattices}
\label{sec:basic}

\subsection{Spherical harmonic oscillator basis \emph{vs.} lattice}
The shell model is a basic concept of nuclear structure~\cite{mayer1955}, and it is crucial for the understanding of key nuclear features~\cite{otsuka2020}. For nuclei that are not too large, it is also a most useful computational tool. The use of the shell model ranges from computations with one or two valence shells above an inert core~\cite{brown1988,caurier2005,shimizu2012,stroberg2019} to the no-core shell-model~\cite{navratil2009,barrett2013}  and other \emph{ab initio} methods that employ a harmonic-oscillator basis as a starting point.

When computing nuclei in large model spaces, however, a shell-model-based approach quickly becomes very expensive. The main reason is that everything that makes the shell model attractive at first glance -- the preservation of symmetries, the well-known analytical properties of the harmonic oscillator basis -- loses its appeal when dealing with several oscillator shells and/or three-nucleon forces. 

The real disadvantage, however, is that the spherical harmonic oscillator basis scrambles the nuclear interactions. The matrix representation of an interaction that is local and short ranged in position space appears non-local and also long ranged with significant off-diagonal matrix elements when expressed in this basis. In pion-less effective field theory, for instance, the three-nucleon force simply is a short-ranged contact interaction. Expressed in the harmonic oscillator basis of the laboratory system, the three-nucleon contact generates tens of terabytes of data in presently employed model spaces~\cite{miyagi2022}. For \emph{ab initio} methods that use this basis as a starting point the generation of the Hamiltonian matrix elements takes more time and storage than the ensuing solution of the quantum many-body problem. In spite of clever approaches~\cite{hebeler2015b,hebeler2021,miyagi2022,miyagi2023EPJA}, generating the matrix elements for the three-body interactions is presently a limiting factor.

Using a lattice for the single-particle basis avoids these problems~\cite{lee2009,lahde2019}. The perceived drawbacks, e.g., the loss of rotational symmetry, are controlled approximations. The lack of degeneracy of states in an angular-momentum multiplet decreases with decreasing lattice spacing. The short range of nuclear interactions then leads to very sparse data objects, further reducing the computational cost for processing and using these interactions in quantum many-body solvers. 

\subsection{Lattice Hamiltonians}
\label{sec:lattice}

Let $D$ be the dimension of the single-particle basis. On a cubic lattice with $L^3$ sites we have 
\begin{equation}
\label{dim}
D=4L^3    
\end{equation}
for nucleons (because of four spin/isospin states per lattice site). Let $\hat{a}^\dagger_{q}$ with 
\begin{equation}
\label{spstate}
    q=(\mathbf{l}, \tau_z,  s_z)
\end{equation}
create a nucleon on the  lattice site $\mathbf{l}=(l_x,l_y,l_z)$, isospin projection $\tau_z$, and spin projection $s_z$. Nuclear Hamiltonians from effective field theories of quantum chromodynamics consist of one-, two-, and three-body operators and have the form 
\begin{align}
\label{ham}
    H &= \sum_{pq} \varepsilon_q^p \hat{a}_p^\dagger \hat{a}_q + {1\over 4}\sum_{pqrs} V^{pq}_{rs} \hat{a}_p^\dagger \hat{a}_q^\dagger \hat{a}_s\hat{a}_r \nonumber \\
    &+ {1\over 36}\sum_{pqrsuv} W^{pqr}_{suv} \hat{a}_p^\dagger \hat{a}_q^\dagger \hat{a}_r^\dagger \hat{a}_v\hat{a}_u\hat{a}_s \ .
\end{align}
The one-body operator is the kinetic energy. In leading order of small lattice spacings, it couples only neighboring sites $p,q$ and therefore consists of ${\cal O}(D)$ terms. Likewise, short-range two- and three-body forces only contain ${\cal O}(D)$ terms because the lattice sites of interacting nucleons must be close to each other. Admittedly, for finite-range interactions the number of nonzero two- and three-body matrix element might be ${\cal O}(D)$ multiplied with a factor of order 30 or so, but that factor is much smaller than $D$ and a constant, i.e., for $L\gg 1$ it is independent of the mass number $A$ and of $L$.

Let us compare this with a harmonic oscillator basis. Equation~(\ref{spstate}) has to be replaced by $q=(n,l,j,j_z,\tau_z)$ where $n$ is the principal quantum number, $l$ the orbital angular momentum, $j$ the total angular momentum, and $j_z$ its projection.  Then the number of nonzero one, two-, and three-body terms in the Hamiltonian~(\ref{ham}) scales approximately as ${\cal O}(D)$, ${\cal O}(D^3)$, and ${\cal O}(D^5)$, respectively. This is so because the conservation of $j_z$ introduces a Kronecker delta and essentially knocks out one index in each of the sums. For $D\approx 4000$ (corresponding to a lattice with $L=10$ or a harmonic oscillator basis with excitations up to single-particle energy $16\hbar\omega$) one would need to generate and process about $10^{18}$ matrix elements for the three-body interaction. For this reason, \emph{ab initio} computations with a harmonic oscillator basis further truncate the matrix elements of the three-nucleon force such that the sum of three excitation quanta stays below about $30\hbar\omega$~\cite{miyagi2022}. In contrast, the lattice computations~\cite{elhatisari2024} face no such problems and work with three-nucleon forces on $L=10$ lattices. 

We note that four-body forces enter in pion-less and in chiral effective field theory at next-to-leading and next-to-next-to-next-to leading order, respectively. For the foreseeable future, it is not practical to work with such forces in the harmonic oscillator basis~\cite{schulz2018}. In contrast, the lattice computations~\cite{epelbaum2010} employed a four-body contact already long ago.  

Thus the do-it-yourself \emph{ab initio} computations use a lattice as the single-particle basis. This exploits the short range of the nuclear force, avoids scrambling of matrix elements, and allows one to use sparse data techniques in the ensuing solutions of the quantum many-body problem. 

We store the nonzero matrix elements of the Hamiltonian~(\ref{ham}) as Python lists
\begin{align}
\label{matlist}
    \varepsilon_q^p & \to \left[\left[p, q, \varepsilon_p^q\right],\ldots\right] \ , \nonumber\\
    V^{pq}_{rs} &\to \left[\left[p,q,r,s,V^{pq}_{rs}\right], \ldots\right] \ , \\
    W^{pqr}_{suv} &\to \left[\left[p,q,r,s,u,v,W^{pqr}_{suv}\right], \ldots\right] \ . \nonumber
\end{align}
Note here that we only store two-body matrix elements $V^{pq}_{rs}$  for $p<q$ and $r<s$ and three-body matrix elements  $W^{pqr}_{suv}$ for $p<q<r$ and $s<u<v$. At present, the code only uses real-valued matrix elements. One could contemplate to further reduce storage needs by exploiting the symmetry of the matrix elements $\varepsilon_q^p=\varepsilon^q_p$ and  $V^{pq}_{rs}= V_{pq}^{rs}$ and  
$W^{pqr}_{suv}=W_{pqr}^{suv}$. However, this would further complicate tensor contractions and only reduce the memory cost by a factor 2 and is therefore not used.

We work on a lattice with spacing $a$ and extent $La$ and use periodic boundary conditions. We denote the nucleon mass as $m$; lattice energies are then given in units of
\begin{equation}
\label{e_lat}
    e_{\rm lat} \equiv \frac{\hbar^2}{2ma^2} \ .
\end{equation}
In this paper we consider a simple nuclear Hamiltonian and choose to work in the framework of pion-less effective field theory at leading order.  The Hamiltonian is 
\begin{align}
\label{hamLO}
    \hat{H} &= \sum_{\mathbf{l} \mathbf{l}'}\sum_{\tau s} T_{\mathbf{l}'}^{\mathbf{l}} \hat{a}_{\mathbf{l}\tau s}^\dagger \hat{a}_{\mathbf{l}'\tau s} \nonumber\\
    &+ {V\over 2}\sum_{\mathbf{l}}\sum_{ss'\tau\tau'} \hat{a}_{\mathbf{l}\tau s}^\dagger \hat{a}_{\mathbf{l}\tau' s'}^\dagger \hat{a}_{\mathbf{l}\tau' s'}\hat{a}_{\mathbf{l}\tau s} \nonumber\\
    &+ W\sum_{\mathbf{l}}\sum_{\tau s} \hat{a}_{\mathbf{l}\tau\uparrow}^\dagger \hat{a}_{\mathbf{l}\tau\downarrow}^\dagger \hat{a}_{\mathbf{l}-\tau s}^\dagger \hat{a}_{\mathbf{l}-\tau s}\hat{a}_{\mathbf{l}\tau \downarrow}\hat{a}_{\mathbf{l}\tau \uparrow}  \ .
\end{align}
The matrix elements of the kinetic energy are the leading-order approximation of the Laplacian on the lattice
\begin{align}
T_{\mathbf{l}'}^{\mathbf{l}} = -\frac{\hbar^2}{2ma^2}\sum_{i = x,y,z}\left(\delta_{\mathbf{l}'}^{\mathbf{l}-\mathbf{e}_i}    -2\delta_{\mathbf{l}'}^{\mathbf{l}} + \delta_{\mathbf{l}'}^{\mathbf{l}+\mathbf{e}_i} \right) .
\end{align}
Here $\mathbf{e}_i$ is a unit vector in the direction $i=x,y,z$. The two-body and three-body contacts are
\begin{equation}
    V = \frac{\hbar^2v}{2ma^2} \quad\mbox{and}\quad W = \frac{\hbar^2w}{2ma^2} \ ,
\end{equation}
where $v$ and $w$ are dimensionless low-energy constants. For three different lattice spacings $a$ we adjusted $v$ and $w$ to  the ground-state energies of $^2$H and $^4$He. As pion-less effective field theory has an uncertainty of about $1/3$ at leading order, approximate adjustments suffice.  Results are shown in Table~\ref{tab:results}. These parameterizations were previously used in Ref.~\cite{rothman2025}. 

\begin{table}
\renewcommand{\arraystretch}{1.2}
\centering
\caption{Calibration of the Hamiltonian. Ground-state energies of $^{3,4}$He (in MeV) from full configuration interaction on lattices of extent $L=4$ and with the lattice spacing $a$ (in fm). The parameters  $v$ and $w$ of the Hamiltonian~(\ref{hamLO}) were adjusted to the ground-state energies of $^2$H and $^4$He.}
\label{tab:results}
\begin{tabular}{cccrrr}
\hline\noalign{\smallskip}
$a$ & $v$    & $w$ & $^4$He   & $^3$He   & $^2$H \\ 
\noalign{\smallskip}\hline\noalign{\smallskip}
2.5 & $-9.0$ & 6.0 & $-29.70$ & $-16.32$ & $-2.48$\\
2.0 & $-8.0$ & 5.5 & $-29.45$ & $-14.84$ & $-2.53$\\
1.7 & $-7.0$ & 4.4 & $-28.97$ & $-10.82$ & $-2.33$\\
\hline\noalign{\smallskip}
\end{tabular}
\end{table}

The Hamiltonian~(\ref{hamLO}) is admittedly probably the simplest  realization of pion-less effective field theory at leading order, see~\cite{kirscher2010,lensky2016,bansal2018} for other implementations. The kinetic energy, two-body interaction, and three-body interaction consist of $28L^3$, $8L^3$, and $4L^3$ terms, respectively. 

We note that the potential of the Hamiltonian~(\ref{hamLO}) is diagonal in the lattice basis. For the three-body interaction this is so because the interaction is on-site and preserves spin projection. Likewise, the two-body potential is diagonal because it also is limited to on-site interactions, preserves spin projection, and exhibits Wigner's SU(4) symmetry. 
As the potential is diagonal in the lattice basis, the Hamiltonian~(\ref{hamLO}) exhibits neither two-particle--two-hole nor three-particle--three-hole excitations. This has the following consequences. The lack of two-particle--two-hole excitations implies that Hartree-Fock computations (which transform the Hamiltonian such that it exhibits no one-particle--one-hole excitations) will yield the bulk of the ground-state energies and that contributions from two-particle--two-hole excitations in coupled-cluster theory are expected to be small. The lack of three-particle--three-hole excitations implies that the normal-ordered two-body truncation is exact (i.e., it is not an approximation) in coupled-cluster with singles and doubles computations~\cite{rothman2025}. We note, finally, that the absence of two-particle--two-hole and three-particle--three-hole excitations in the Hamiltonian does not imply that the two-particle--two-hole amplitudes vanish in coupled-cluster theory; it only explains that they are small.

\subsection{Full configuration interaction on lattices}
\label{sec{fci}}
Full configuration interaction (FCI) is available for nuclei with mass numbers $A=2,3,4$. We construct an $A$-body basis and use the lists of matrix elements~(\ref{matlist}) to construct the matrices for the kinetic energy, the two-body interaction, and the three-body interaction. These are stored  as a \texttt{\detokenize{csr_matrix}} in the compressed sparse row format of the Python module \texttt{scipy.sparse}. This format is efficient for various operations, and the lowest few eigenvalues and eigenstates are computed with \texttt{linalg.eigsh} from \texttt{scipy.sparse}.  The results presented in Table~\ref{tab:results} were computed with full configuration interaction. On lattices with $L=4$, the $^4$He nucleus has a Hamiltonian matrix with a dimension of $(64)^4 + 2\binom{64}{2}\approx 25\times 10^6$ (counting states with zero spin projection) and the required storage exceeds the 20~GB that are typically available on laptops. For our special Hamiltonian from leading-order pion-less effective field theory, however, the matrix dimension can be reduced to $64^{4}\approx 17\times 10^6$ because one can exclude configurations where both protons have the same spins (and both neutrons have the opposite same spins).

\subsection{Hartree Fock on the lattice}
\label{sec:HF}

The Hartree-Fock (HF) computations start from a one-body density matrix $\rho^p_q$ where $A$ nucleons occupy (or partly occupy) a set of lattice sites. The Hartree-Fock Hamiltonian has matrix elements
\begin{equation}
\label{HFham}
    \left({H_{\rm HF}}\right)^p_q = \varepsilon^p_q + \sum_{sr}V^{pr}_{qs}\rho^s_r  + {1\over 2} \sum_{rsuv}W^{pru}_{qsv} \rho^s_r \rho^v_u \ . 
\end{equation}
The contractions with the density are performed such that the sums in Eq.~(\ref{HFham}) are replaced by loops over the nonzero matrix elements~(\ref{matlist}). For example, the second term on the right-hand side of Eq.~(\ref{HFham}) is computed by looping over the list of non-zero matrix elements $V^{pr}_{qs}$.   
\begin{lstlisting}[language=Python, caption=Python example of sparse data operations in Hartree Fock, label=listing1]
    import numpy as np
    result = np.zeros_like(density)
    for element in v2List:
        [p,r,q,s,val] = element
        result[p,q] += val*density[r,s]
        result[r,q] -= val*density[p,s]
        result[p,s] -= val*density[r,q]
        result[r,s] += val*density[p,q]
\end{lstlisting}
We remind the reader that the lists~(\ref{matlist}) only store two-body matrix elements with $(p<q)$ and $(r<s)$. As the sums in Eq.~(\ref{HFham}) are unrestricted one needs  to sum over permutations of $(p,q)$ and $(r,s)$. Thus the implementation of the last term on the right-hand side of Eq.~(\ref{HFham}) then requires 36 permutations of three-body matrix elements.  

The  Hartree-Fock energy  
\begin{equation}
    E_{\rm HF} = \sum_{pq}\varepsilon^p_q\rho_p^q + {1\over 2}\sum_{sr}V^{pr}_{qs}\rho_p^q\rho^s_r  + {1\over 6} \sum_{qruv}W^{pru}_{qsv}\rho_p^q \rho^s_r \rho^v_u  
\end{equation}
is computed in a similarly efficient way. In the iterations of the Hartree-Fock equations, the convergence criterion is based on the convergence of the density matrix that enters the Hartree-Fock Hamiltonian~(\ref{HFham}).

\subsection{Coupled-cluster computations on lattices}
\label{sec:ccm}

The nuclear coupled-cluster cluster method was reviewed in Refs.~\cite{hagen2014,hergert2020}. Here we recall essential aspects and point out unique features when performing these calculations on the lattice.

In the harmonic-oscillator basis, coupled-cluster computations would start from a Hartree-Fock reference state. On a lattice, however, this is probably not advised, because the Hartree-Fock basis scrambles matrix elements and takes away the advantage of sparse interaction input brought on by using a lattice. This means that the impact of singles amplitudes will be significant. 

The reference state is
\begin{equation}
\label{ref}
    |\phi\rangle = \prod_{i=1}^A\hat{a}^\dagger_i|0\rangle \ .
\end{equation}
Here and in what follows we use the convention that indices $i, j, k, \ldots$ refer to occupied (hole) states and $a,b,c,\ldots$ to unoccupied (particle) states. If no distinction is made, we use indices $p,q,r,\ldots$. Coupled-cluster computations then start from the normal-ordered Hamiltonian
\begin{align}
\label{HamNO}
    \hat{H} &= E_{\rm ref} + \sum_{pq} F_q^p\left\{\hat{a}_p^\dagger \hat{a}_q \right\}+ {1\over 4}\sum_{pqrs}\Gamma^{pq}_{rs}\left\{\hat{a}_p^\dagger \hat{a}_q^\dagger \hat{a}_s \hat{a}_r\right\} \nonumber\\
    &+{1\over 36}\sum_{pqrstu}W^{pqr}_{stu}\left\{\hat{a}_p^\dagger \hat{a}_q^\dagger \hat{a}_r^\dagger \hat{a}_u\hat{a}_t\hat{a}_s\right\} \ ,
\end{align}
where the brackets $\{\cdot\}$ indicate normal ordering with respect to the reference state. In Eq.~(\ref{HamNO}) we introduced the reference-state energy and normal-ordered matrix elements
\begin{align}
\label{NO-matele}
E_{\rm ref} &=\sum_i \varepsilon_i^i + {1\over 2}\sum_{ij}V_{ij}^{ij} + {1\over 6}\sum_{ijk}W_{ijk}^{ijk} \ , \nonumber\\ 
F_q^p &= \varepsilon_q^p + \sum_i V_{iq}^{ip} +{1\over 2} \sum_{ij}W^{ijp}_{ijq} \ ,  \\
\Gamma^{pq}_{rs} &= V^{pq}_{rs} + \sum_i W_{irs}^{ipq} \ . \nonumber
\end{align}
We note that the normal ordering is performed with respect to reference states that break rotation and translation invariance. In the shell model, one can devise a normal ordering that respects rotational symmetry~\cite{Frosini:2021tuj}. An attempt to do this here would probably also lead to a loss of sparsity.    

In coupled-cluster theory one computes the matrix elements of the similarity transformed Hamiltonian
\begin{equation}
\label{Hsim}
    \overline{H}\equiv e^{-\hat{T}}\hat{H} e^{\hat{T}} \ .
\end{equation}
In the coupled-cluster with singles and doubles (CCSD) approximation, the cluster operator is
\begin{align}
\label{cluster}
    \hat{T} = \sum_{ia}t_i^a\hat{a}^\dagger_a\hat{a}_i + 
    {1\over 4} \sum_{ijab}t_{ij}^{ab}\hat{a}^\dagger_a\hat{a}^\dagger_b\hat{a}_j\hat{a}_i \ .
\end{align}
It consists of one-particle--one-hole and two-particle--two-hole excitations. 
The amplitudes $t_i^a$ and $t_{ij}^{ab}$ are determined by solving the nonlinear set of equations
\begin{align}
\label{ccsd}
    \overline{H}_i^a&\equiv \langle\phi|\hat{a}^\dagger_i\hat{a}_a \overline{H}|\phi\rangle = 0 \ , \nonumber\\
    \overline{H}_{ij}^{ab}&\equiv \langle\phi|\hat{a}^\dagger_i\hat{a}^\dagger_j\hat{a}_b\hat{a}_a \overline{H}|\phi\rangle = 0 \ .
\end{align}

We refer to the left-hand sides of these equations as the coupled-cluster residuals and denote them by $r$. In what follows we use an equally compact notation and refer to $t^a_i$ and $t^{ab}_{ij}$ simply as $t$.
The CCSD Eqs.~(\ref{ccsd}) are usually solved using the inexact Newton's method~\cite{yang2020}. This is an iterative approach which updates the coupled cluster amplitudes at step $(k+1)$  as  
\begin{align}
\label{iter}
    t^{(k+1)} &\to t^{(k)} - \left[ J^{(k)} \right] ^{-1} r^{(k)} \ . 
\end{align}
Here, $J^{(k)}$ is the coupled-cluster Jacobian, which, in its simplest approximation, consists of the diagonal parts $F_p^p$ of the Fock matrix, i.e. $F_i^i-F_a^a$ for the singles amplitudes and $F_i^i+F_j^j-F_a^a-F_b^b$ for the doubles amplitudes~\cite{kjonstad2020}. For product states on the lattice, however, these energy differences may vanish. This is because of translational invariance; the reference state is not unique but rather highly degenerate. To avoid this problem, we add a positive constant to the difference in the first iteration. In subsequent iterations we replace the diagonal matrix elements of the Fock matrix by the diagonal elements $\overline{H}_p^p$ of the similarity-transformed Hamiltonian~(\ref{Hsim}). This is a more accurate approximation of the coupled-cluster Jacobian.

Our solver of the coupled-cluster equations also allows one to mix old and new solutions via
\begin{equation}
    t^{(k+1)}\leftarrow a t^{(k+1)} +(1-a)t^{(k)} \ ,
\end{equation}
(where $0\le a\le 1$), and we use the direct inversion of the iterative subspace (DIIS) method~\cite{pulay1980,scuseria1986} for an accelerated convergence. 

Once the coupled-cluster amplitudes $t_i^a$ and $t_{ij}^{ab}$ are known, the ground-state energy becomes
\begin{equation}
\label{Eccs}
    E_{\rm CCS} = E_{\rm ref} + \sum_{ia}F_a^it_i^a + \frac{1}{2}\sum_{ijab}\Gamma_{ab}^{ij}t_i^at_j^b
\end{equation}
in the CCS approximation, and 
\begin{equation}
\label{Eccsd}
    E_{\rm CCSD} = E_{\rm CCS} + \frac{1}{4}\sum_{ijab}\Gamma_{ab}^{ij}t_{ij}^{ab}
\end{equation}
for CCSD. Equations~(\ref{Eccs}) and (\ref{Eccsd}) are both valid in the normal-ordered two-body approximation, where one truncates the 
normal-ordered Hamiltonian~(\ref{HamNO}) at two-body rank. The energy correction from three-body forces was presented in Ref.~\cite{hagen2007a}.  

In \texttt{NuLattice} the coupled cluster computations start by construction of the normal-ordered two-body Hamiltonian~(\ref{HamNO}), truncated at two-body rank. Coupled-cluster theory distinguishes between particle and hole states, and the normal-ordered two-body Hamiltonian contains six types of such matrix elements, namely $\Gamma^{ab}_{cd}$, $\Gamma^{ab}_{ci}$, $\Gamma^{ab}_{ij}$, $\Gamma^{ai}_{bj}$, $\Gamma^{ai}_{kl}$, and $\Gamma^{ij}_{kl}$ (here we exploited that the Hamiltonian is Hermitian). In general, the particle indices run over many more single-particle states than the hole indices. Thus we store $\Gamma^{ab}_{cd}$ and $\Gamma^{ab}_{ci}$ as sparse lists~(\ref{matlist}) but $\Gamma^{ab}_{ij}$, $\Gamma^{ai}_{bj}$, $\Gamma^{ai}_{kl}$, and $\Gamma^{ij}_{kl}$ as dense four-dimensional \texttt{numpy} arrays. Likewise, we store the matrix elements $\overline{H}^{ab}_{ij}$ and the coupled-cluster amplitudes $t^{ab}_{ij}$ as dense four-dimensional \texttt{numpy} arrays. The storage demands of these \texttt{numpy} arrays sets the limits on what lattice sizes and nuclei can be held in fast memory and on what can thus be computed efficiently.  
We note here that the lion's share of the required memory comes from storing about ten amplitudes $t^{ab}_{ij}$ when using the DIIS method. 

There are two different types of contractions done by the code, sparse-dense and dense-dense contractions. Both lead to dense \texttt{numpy} arrays, and this makes the coupled-cluster method simple to implement and efficient. The sparse-dense contractions involve the sparse objects $\Gamma^{ab}_{cd}$ or $\Gamma^{ab}_{ci}$ and dense \texttt{numpy} arrays $t_i^a$ and/or $t_{ij}^{ab}$. For example, the contraction 
\begin{equation}
	\overline{H}_i^a \leftarrow -{1\over 2} \sum_{cdk} \Gamma_{cd}^{ak}t^{cd}_{ki}\ ,
\end{equation}
which occurs in calculating $\overline{H}_i^a$, is handled  by the following code, where we loop over the the sparse list of the nonzero elements $\Gamma_{cd}^{ak}$.

\begin{lstlisting}[language=Python, 
caption=Python example of a sparse-dense tensor contraction that yields a dense tensor., 
label=listingSparseContract]
import numpy as np
result = np.zeros((pnum, hnum))
for vals in v_ppph:
    c,d,a,k,v = vals
    result[a,:] -= 0.5*v*t2[c,d,k,:]
\end{lstlisting}

The dense-dense contractions involve tensors that are stored as \texttt{numpy} arrays. They are performed by \texttt{\detokenize{opt_einsum}}~\cite{Smith2018}, which provides the same interface as \texttt{numpy.einsum} with optimized evaluation strategies. The contraction
\begin{equation}
	\overline{H}_{ij}^{ab} \leftarrow  {1\over 4}\sum_{cdkl}\Gamma^{kl}_{cd}t^{cd}_{ij}t^{ab}_{kl}\ ,
\end{equation}
for example, is performed as follows.
\begin{lstlisting}[language=Python, caption=Python example of a dense-dense tensor contraction that yields a dense tensor., label=listingDenseContract]
from opt_einsum import contract
result = 0.25*contract("cdkl,cdij,abkl->abij",
                       v_pphh,t2,t2,
                       optimize="greedy")
\end{lstlisting}
Here, the function \texttt{contract} from \texttt{\detokenize{opt_einsum}} automatically factorizes the full contraction into several sequential contractions involving intermediate tensors. These examples show that it is straightforward to translate coupled-cluster expressions into Python code. Ultimately, coupled-cluster computations using lattices with $L=5$ are possible on laptops  because no dense tensors with more than two particle indices are used or created in such calculations.  

Table~\ref{tab:CCSDresources} lists the resource requirements for coupled-cluster computations of $^{16}$O on lattices with a spacing $a=2.5$~fm at size $L$. The number of iterations of the coupled-cluster Eqs.~(\ref{ccsd}) is also given. We see that the memory required to store the doubles amplitudes $t_{ij}^{ab}$ roughly increases as $L^6$, because the number of single-particle states (and the number of particle states) scales as $L^3$, see Eq.~(\ref{dim}).

\begin{table}
\setlength{\tabcolsep}{5pt} 
\renewcommand{\arraystretch}{1.2}
\centering
\caption{Memory (in MB), total run time (in seconds), and number of iterations used in CCSD computations of $^{16}$O on lattice with size $L$ and spacing $a=2.5$~fm. Sparse matrices were used and ten sets of amplitudes were stored in the 
direct inversion of the iterative subspace (DIIS) method. The system used was a 12 core Intel i7-9750H CPU @ 2.60 GHz with 32 GB of RAM.}
\label{tab:CCSDresources}
\begin{tabular}{rrcrc}
\hline\noalign{\smallskip}
$L$	&	Memory 	& Memory$/L^6$ 	& Time	&	Iterations \\
\noalign{\smallskip}\hline\noalign{\smallskip}
3	&	600		&	0.82		&	34	&	32 \\
4	&	4,100	&	1.00		&	330 &	27 \\
5	&	15,900	&	1.02		&  1,517	&	23 \\
\hline\noalign{\smallskip}
\end{tabular}
\end{table}

Let us also discuss the runtime requirements. Using dense tensors, the most expensive CCSD diagram (in the-normal-ordered two-body approximation) has a cost of ${\cal O}(A^2D^4)$ for a nucleus of mass number $A$ and a lattice with $D$ single-particle states~(\ref{dim}). Here we assumed $A\ll D$. In the sparse formulation, the cost of that diagram reduces to ${\cal O}(A^2N_{\rm nz})$ where $N_{\rm nz}$ is the number of nonzero normal-ordered two-body matrix elements. We have $N_{\rm nz}={\cal O}(D)$ in pion-less effective field theory and $N_{\rm nz}={\cal O}(D^2)$ for local long-range forces. Thus, the usage of sparse tensors matters a lot in the coupled-cluster computations. Our implementation of CCSD also uses dense tensors for several parts of the Hamiltonian. The time to compute the most costly diagram involving dense tensors is from a contraction of two doubles amplitudes with $\Gamma_{ab}^{ij}$ and asymptotically scales as ${\cal O}(A^3 D^3)$. If we used a sparse implementation of $\Gamma_{ab}^{ij}$, that cost would be ${\cal O}(A^2D^2 N_{\rm nz})$, which is probably about similar. Given the efficient implementation of the dense tensor contractions via  \texttt{\detokenize{opt_einsum}} a sparse implementation of $\Gamma_{ab}^{ij}$ is probably not urgent.

We finally note that the largest computations reported in Table~\ref{tab:CCSDresources} use lattices with size $L=5$ and thus involve $D=500$ single-particle states. Let us compare this to a harmonic oscillator basis with oscillator spacing $\hbar\omega$. There, the number of single-particle states up to and including the energy $(N_{\rm max}+3/2)\hbar\omega$ is $D_{\rm ho}=(2/3)(N_{\rm max}+1)(N_{\rm max}+2)(N_{\rm max}+3)$.  For $N_{\rm max}=7$ we have $D_{\rm ho}=480$ which thus is comparable to a lattice with $L=5$. It is clear that \texttt{NuLattice} allows one to run non-trivial computations on a laptop or desktop.    

\subsection{In-medium similarity renormalization group on the lattice}

The IMSRG~\cite{tsukiyama2011} was reviewed in Refs.~\cite{hergert2016, stroberg2019}, so here we only recap the most essential details necessary to understand the IMSRG calculations in this work.

As with coupled-cluster theory, one starts from a Slater determinant reference state, Eq.~\eqref{ref}, and the normal-ordered Hamiltonian, Eq.~\eqref{HamNO}.
To solve the many-body Schr\"odinger equation, the IMSRG computes a continuous unitary similarity transformation of the Hamiltonian
\begin{equation}
\label{HsimIMSRG}
    \overline{H}(s) \equiv U(s)\hat{H} U^\dagger(s) 
\end{equation}
via the integration of the IMSRG flow equation
\begin{equation}
    \label{IMSRGflow}
    \frac{d\overline{H}(s)}{ds} = \left[\eta(s), \overline{H}(s)\right]
\end{equation}
with the initial condition that $U(s=0) = 1$.
The generator $\eta(s)$ is chosen to produce the desired unitary transformation.

In general, even if one starts with a Hamiltonian with only up to normal-ordered two-body terms,
\begin{align}
    \overline{H}(s=0) &= E_{\rm ref} + \sum_{pq} F_q^p\left\{\hat{a}_p^\dagger \hat{a}_q \right\} + {1\over 4}\sum_{pqrs}\Gamma^{pq}_{rs}\left\{\hat{a}_p^\dagger \hat{a}_q^\dagger \hat{a}_s \hat{a}_r\right\} \ ,
\end{align}
the right-hand side of Eq.~\eqref{IMSRGflow} will contain three-body (and later on higher-body) terms.
This means that the IMSRG induces effective many-body interactions,
\begin{align}
    \overline{H}(s) &= E(s) + \sum_{pq} F_q^p(s)\left\{\hat{a}_p^\dagger \hat{a}_q \right\}+ {1\over 4}\sum_{pqrs}\Gamma^{pq}_{rs}(s)\left\{\hat{a}_p^\dagger \hat{a}_q^\dagger \hat{a}_s \hat{a}_r\right\} \nonumber\\
    &+{1\over 36}\sum_{pqrstu}W^{pqr}_{stu}(s)\left\{\hat{a}_p^\dagger \hat{a}_q^\dagger \hat{a}_r^\dagger \hat{a}_u\hat{a}_t\hat{a}_s\right\} + \dots\ .
\end{align}
The IMSRG(2) approximation truncates the IMSRG equation at the normal-ordered two-body level for all $s$, discarding, e.g., the induced normal-ordered three-body interactions $W^{pqr}_{stu}(s)$.
Recent work has extended the IMSRG to the IMSRG(3) level, where this truncation is relaxed, and shown the IMSRG(2) to be accurate for ground-state properties of medium-mass nuclei~\cite{heinz2021,Stroberg2024,Heinz2025}.
In this work, we exclusively study the IMSRG(2) approximation.

In the IMSRG(2), the generator $\eta(s)$ is chosen such that
\begin{align}
    F^i_a(s\to\infty) &= 0\ ,\nonumber \\
    \Gamma^{ij}_{ab}(s\to\infty) & = 0\ ,
\end{align}
reaching the same decoupling condition as in CCSD:
\begin{align}
    \langle\phi|\hat{a}^\dagger_i\hat{a}_a \overline{H}(s\to\infty)|\phi\rangle &= 0 \ , \nonumber\\ 
    \langle\phi|\hat{a}^\dagger_i\hat{a}^\dagger_j\hat{a}_b\hat{a}_a \overline{H}(s\to\infty)|\phi\rangle &= 0 \ .
\end{align}
We note, however, that the IMSRG keeps $\overline{H}$ Hermitian and therefore also decouples one-particle--one-hole and two-particle--two-hole de-excitations. Various choices for generators exist (see Ref.~\cite{hergert2016} for details), and all are equivalent up to errors induced by the IMSRG(2) approximation.
The final IMSRG energy is then simply
$E_\mathrm{IMSRG} = E(s\to\infty)$.

The main computational task in the IMSRG is the evaluation of the right-hand side of Eq.~\eqref{IMSRGflow}.
The computational cost is dominated by the normal-ordered two-body matrix elements resulting from the commutator of the two-body part of the generator, with matrix elements $\eta^{pq}_{rs}$, and the two-body part of the Hamiltonian, with matrix elements $\Gamma^{pq}_{rs}$.
One such contribution is
\begin{equation}
    X^{pq}_{rs}\leftarrow \frac{1}{2}\sum_{tu}(\bar{n}_t\bar{n}_u - n_t n_u) \left(\eta^{pq}_{tu} \Gamma^{tu}_{rs} - \Gamma^{pq}_{tu} \eta^{tu}_{rs}\right)\ ,
\end{equation}
with the occupation number $n_p = 1$ for hole states, 0 for particle states, and $\bar{n}_p = 1 - n_p$.
This can be easily evaluated in Python using \texttt{\detokenize{opt_einsum}}. 
\begin{lstlisting}[language=Python, caption=Python example of dense IMSRG(2) contraction, label=listingIMSRGContract]
from opt_einsum import contract
result = 0.0
result += contract(
    "pqtu,turs,t,u->pqrs", 
    eta,gamma,nbar,nbar, 
    optimize="greedy",
)
result -= contract(
    "pqtu,turs,t,u->pqrs", 
    eta,gamma,n,n, 
    optimize="greedy",
)
result -= contract(
    "pqtu,turs,t,u->pqrs", 
    gamma,eta,nbar,nbar, 
    optimize="greedy",
)
result += contract(
    "pqtu,turs,t,u->pqrs",
    gamma,eta,n,n, 
    optimize="greedy",
)
result *= 0.5
\end{lstlisting}

As the Hamiltonian $\overline{H}(s)$ is transformed and evolved, its structure changes.
As a result, it is not as straightforward to develop a simple sparse approach to solving the IMSRG equations, as the sparsity and the sparsity structure is not constant.
For this reason we do not explore sparse IMSRG(2) calculations in this work,
focusing on lattice sizes $L=2,3$ where the dense approach above is computationally tractable.
At $L=3$, the storage cost of $\Gamma^{pq}_{rs}$ is around 1~GB, and the total IMSRG(2) solution consumes a peak memory of about 20~GB due to the storage of multiple copies for the fifth-order Runge-Kutta solver used and of intermediates in the tensor contractions.
Such memory costs and the associated computational costs prevent IMSRG(2) calculations from being performed on larger lattices.
However, past work has explored importance truncation in IMSRG calculations~\cite{Hoppe2022}, finding significant room for compression in IMSRG(2) calculations.
Such developments can be leveraged to also allow for sparse IMSRG calculations on a lattice, which would enable calculations on larger lattices. Similar efforts in self-consistent Green's function theory may also prove instrumental when adding those methods to \texttt{NuLattice}~\cite{Porro:2021rjp}.

\section{Program package}
\label{sec:package}
The package consists of several Python modules and IPython Notebooks. Each function is a pure function. Documentation of each function is created using reStructuredText via Sphinx~\cite{sphinx}. Driver routines exist to perform computations. In this Section we briefly summarize the modules of the package.

\subsection{Module \texttt{constants.py}}
\label{mod:const}
This module provides the user with basic constants needed for computations. 

\subsection{Module \texttt{lattice.py}}
\label{mod:lattice}

This module contains routines that create the lattice and nuclear interactions from pion-less effective field theory. The basis consists of a list of single-particle states~(\ref{spstate}).  Matrix elements of operators are lists~(\ref{matlist}) of nonzero elements.

\subsection{Module \texttt{references.py}}
\label{mod:refs}

This module provides the user with reference states that are used in the modules \texttt{HF}, \texttt{CCM}, and \texttt{IMSRG}. 

\subsection{Module \texttt{\detokenize{FCI}}}
\label{mod:exact}

This module contains routines for full configuration interaction computations of nuclei with mass numbers $A=2,3,4$.

\subsection{Module \texttt{\detokenize{HF}}}
\label{mod:hf}
This module contains the routines for Hartree-Fock computations of nuclei.

\subsection{Module \texttt{IMSRG}}
\label{mod:imsrg}

This module contains the routines for IMSRG(2) computations.

\subsection{Module \texttt{CCM}}
\label{mod:ccm}

This module contains the routines for the coupled-cluster with singles and doubles (CCSD) computations.

\subsection{Examples}
\label{sec:example}
The following programs illustrate the usage of the package. They are available as IPython notebooks and Python files. 

\begin{description}
\item \texttt{\detokenize{Example_FCI.ipynb}} performs full configuration interaction computations (exact diagonalizations) of the nuclei $^2$H and $^{3,4}$He. \\

\item \texttt{\detokenize{Example_Hartree_Fock.ipynb}} performs a Hartree-Fock computation of the nucleus $^{16}$O. \\

\item \texttt{\detokenize{Example_CCSD.ipynb}} and \texttt{\detokenize{Example_CCSD2.ipynb}} perform a coupled-cluster computation (in the CCSD approximation) of the nucleus $^{16}$O. \\

\item \texttt{\detokenize{Example_IMSRG.ipynb}} performs an IMSRG(2) computation of the nucleus $^{3}$He. \\

\item The directory \texttt{\detokenize{reproduction_arxiv_2509.08771}} contains additional example programs to compute some of the values in Fig.~\ref{fig:LSize} and Table~\ref{tab:IMSRGresults}.
\end{description}

\subsection{Benchmarks}
We performed several benchmark computations and recorded those results.

\section{Results}
\label{sec:results}

In what follows we compute ground-state energies for several nuclei using a leading-order Hamiltonian from pion-less effective field theory. We will see that the convergence pattern of coupled-cluster computations on the lattice is different than from what is known in the harmonic oscillator basis (at least for the employed Hamiltonian).  As we will see, one-particle--one-hole excitations play an outsized role in certain nuclei. 

\subsection{Results for light nuclei with $A=2, 3,4$ }
\label{sec:benchmark}

\begin{figure}[htb]
\centering
\includegraphics[width=1.0\columnwidth]{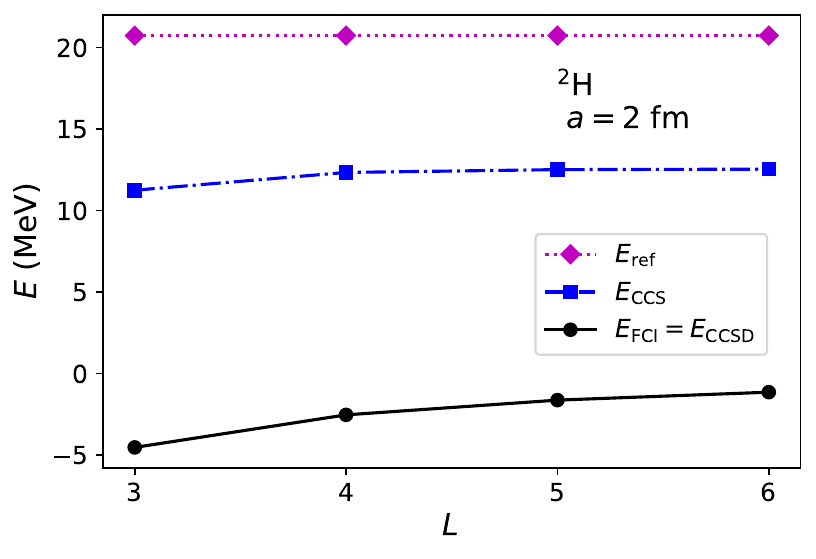}
\caption{Ground-state energy of $^2$H as a function of the lattice size $L$ for a spacing $a=2$~fm, computed via coupled-cluster with singles (CCS) and coupled-cluster with singles and doubles (CCSD), which yields the same result as full configuration interaction (FCI). The energy of the reference state is $E_{\rm ref}$.}\label{fig:H2vsL}
\end{figure}

We start with computations of the deuteron and show ground-state energies as a function of the lattice size $L$ in Fig.~\ref{fig:H2vsL}. The coupled-cluster with singles (CCS) approximation only yields a smaller contribution to the ground-state energy, while coupled-cluster with singles and doubles (CCSD)  yields the lion's share (and is indeed the exact solution of the two-body system).  We were unable to obtain converged Hartree-Fock solutions for the deuteron. It is not easy to understand why that is so. On the one hand, we could perform the CSS computations and one might expect that CCS is in some sense close to Hartree Fock. On the other hand, the concept of a mean field is not adequate for a two-body system, and the CCS computation indicates that the deuteron is not bound when only one-particle--one-hole excitations are included.

Figure~\ref{fig:H2vsL} also displays the known convergence patterns of bound-state energies on lattices with periodic boundary conditions~\cite{luscher1985,konig2017}. The periodic boundary conditions allow for tunneling processes (which lower the kinetic energy) to take place. Such energy gains are exponential in the dimensionless product $k_{\rm sep}aL$, where $k_{\rm sep}$ is a separation momentum (and $aL$ the spatial extent of the lattice). For the weakly bound deuteron, the separation momentum is relatively small and one needs correspondingly large lattices. 
In what follows we will see that this is different for nuclei that are more strongly bound.   

We turn to nuclei with mass numbers $A=3,4$. Table~\ref{tab:benchmarksL4} shows the ground-state energies of $^{3,4}$He for a lattice with extent $L=4$ and spacing $a$ computed with Hartree Fock (HF), coupled-cluster with singles (CCS), coupled-cluster with singles and doubles (CCSD), and full configuration interaction (FCI). The energy of the reference state is denoted as $E_{\rm ref}$. The interaction parameters are as in Table~\ref{tab:results}. 

\begin{table}
\setlength{\tabcolsep}{4pt}
\centering
\caption{Ground-state energies (in MeV) of $^{3,4}$He for a lattice with extent $L=4$ and spacing $a$ (in fm), computed with Hartree Fock (HF), coupled-cluster with singles (CCS), coupled-cluster with singles and doubles (CCSD), and full configuration interaction (FCI). The energy of the reference state is denoted as $E_{\rm ref}$. }
\label{tab:benchmarksL4}
\begin{tabular}{c|rrr|rrr}
\hline
\multicolumn{1}{c|}{} & \multicolumn{3}{c|}{$^3$He} & \multicolumn{3}{c}{$^4$He}\\
\hline
 $a$ & 1.7  & 2.0 & 2.5 & 1.7 & 2.0 & 2.5\\
$E_{\rm ref}$&  $10.04$ &  $-2.59$ &  $-9.95$ &  $-2.87$ & $-10.37$ & $-19.91$ \\
HF           &  $-4.62$ & $-12.09$ & $-15.18$ & $-26.37$ & $-27.65$ & $-29.06$ \\
CCS          &  $-4.09$ & $-11.85$ & $-15.08$ & $-26.82$ & $-28.52$ & $-29.28$ \\
CCSD         &  $-4.80$ & $-12.18$ & $-15.22$ & $-26.41$ & $-27.73$ & $-29.11$ \\
FCI          & $-10.82$ & $-14.84$ & $-16.32$ & $-28.97$ & $-29.45$ & $-29.70$ \\
\hline\noalign{\smallskip}
\end{tabular}
\end{table}

Several features stand out. 
First, the results from CCS and CCSD are close to those from HF, and CCSD yields little change in energy compared to CCS. Second, the CCSD results are closer to the FCI benchmarks than HF, although the energy gain is relatively small. Third, the CCS results for $^4$He are lower in energy than those from CCSD.

These results suggest that the lattice computations of nuclei are different than those in the harmonic oscillator basis, at least for the local (on-site) interactions used in this work. \citet{bansal2018} presented CCSD computations with interactions from pion-less effective field theory. In contrast to the present work, those authors used the harmonic oscillator basis, employed smeared out and non-local contact interactions, and worked with pion-less effective field theory at  next-to-leading order.  Their CCSD results were close to FCI benchmarks.    

We note that CCS results in Table~\ref{tab:benchmarksL4} are close to, but different from, the HF results. This is no surprise. The CCS approximation is equivalent to Hartree Fock only in the Hartree-Fock basis (where singles fulfill $T_1=0$). In all other bases, CCS yields a non-Hermitian similarity transformed Hamiltonians whose ground-state energy is computed via Eq.~(\ref{Eccs}).    

\begin{figure}[htb]
\centering
\includegraphics[width=1.0\columnwidth]{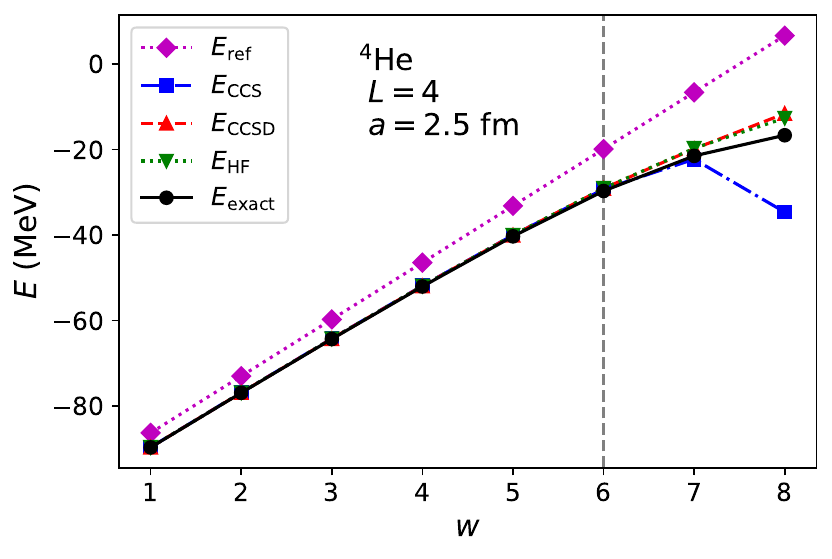}
\caption{Ground-state energy of $^4$He on a lattice with $L=4$ and $a=2.5$~fm, as a function of the three-body strength $w$, from coupled cluster with singles (CCS), coupled cluster with singles and doubles (CCSD), Hartree Fock (HF), and compared to full configuration interaction (FCI). The energy of the reference state is $E_{\rm ref}$. The vertical dashed line marks the physical point.}\label{fig:He4ergs}
\end{figure}

It is instructive to study ground-state energies for different strengths $w$ of the repulsive three-body contact. Figure~\ref{fig:He4ergs} shows the ground-state energy of $^4$He on a lattice with $L=4$ and $a=2.5$~fm. On the admittedly large scale of the plot (spanning about 100~MeV) all computed energies are close to each other and significantly improve upon the energy $E_{\rm ref}$ of the lattice reference state. The vertical dashed line marks the ``physical point'' as listed in Table~\ref{tab:results}. We also see that the CCS results become unphysical for too large strengths $w$ (as they are below the FCI results). This is not entirely surprising: Coupled-cluster theory is based on the non-Hermitian Hamiltonian~(\ref{Hsim}) and therefore is not a variational method. Inspection shows that this breakdown happens because the hole states of the Fock matrix $F^p_q$ in Eq.~(\ref{NO-matele}) become higher in energy than the particle states. (We also note that the energy of the reference state becomes unbound.)   

\begin{figure}[htb]
\centering
\includegraphics[width=1.0\columnwidth]{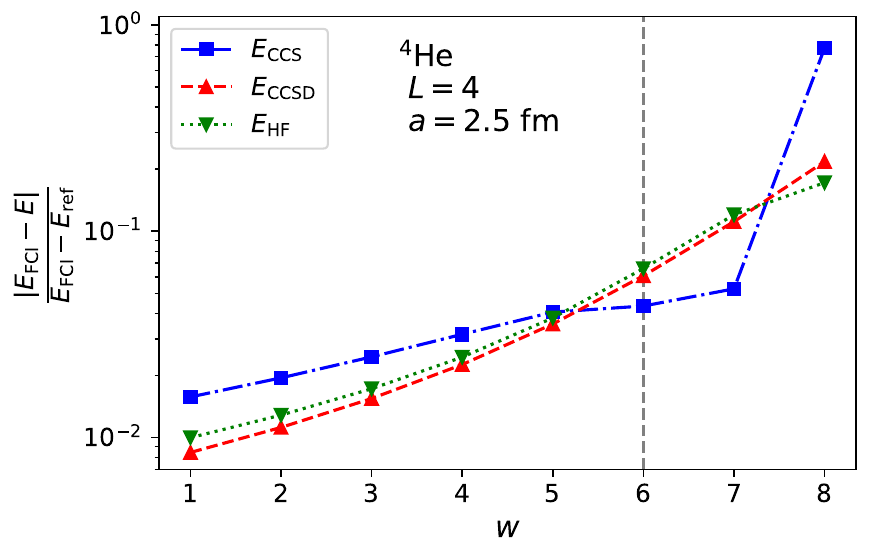}
\caption{Absolute ratio of the missing  energy $E_{\rm FCI}-E$ and the correlation energy  $E_{\rm FCI}-E_{\rm ref}$  in computations of $^4$He on a lattice with $L=4$ and $a=2.5$~fm, as a function of the three-body strengths $w$, from coupled cluster with singles (CCS), coupled cluster with singles and doubles (CCSD), and Hartree Fock (HF). The vertical dashed line marks the physical point.}\label{fig:He4miss}
\end{figure}

Figure~\ref{fig:He4miss} shows the absolute ratio of the missing energy $E_{\rm FCI}-E$ and the correlation energy  $E_{\rm FCI}-E_{\rm ref}$ for the HF, CCS, and CCSD computations. For sufficiently small three-body strengths, the CCSD results are more accurate than HF and CCS. This changes only around $w\approx 6$ (where CCS breaks down and incidentally becomes more accurate) and for very large couplings (where also CCSD starts to break down and becomes less accurate than HF).    

To better understand the lattice results, we also compare with IMSRG(2) and with CCSD computations that use the Hartree-Fock state as a reference. We limit these computations to lattices with $L=3$ because IMSRG(2) does not exploit the sparsity of the lattice interaction, and the CCSD computations employ the dense interaction in the Hartree-Fock basis. Results are shown in Table~\ref{tab:benchmarksL3}.

\begin{table}
\setlength{\tabcolsep}{4pt}
\centering
\caption{Ground-state energies (in MeV) of $^{4}$He for a lattice with extent $L=3$ and spacing $a=2$~fm, computed with Hartree Fock (HF), coupled-cluster with singles and doubles (CCSD), the in-medium similarity renormalization group (IMSRG), and full configuration interaction (FCI). For CCSD and IMSRG, we consider both reference states constructed from the lattice basis and the HF basis. }
\label{tab:benchmarksL3}
\begin{tabular}{ccccccc}
\hline\noalign{\smallskip}
\multirow{2}{*}{Nucleus} & \multirow{2}{*}{HF} & \multicolumn{2}{c}{CCSD} & \multicolumn{2}{c}{IMSRG(2)}
& \multirow{2}{*}{FCI}\\
 & & lattice & HF & lattice & HF & \\
\hline\noalign{\smallskip}
$^4$He  & $-29.79$ & $-29.83$ & $-30.15$ &  $-29.14$  & $-30.15$  & $-32.99$ \\
\hline
\end{tabular}
\end{table}

We see that the CCSD and IMSRG(2) results are close to each other, that the difference for the lattice or HF reference state is small (compared to the energy difference between HF and FCI), and that they  add little correlation energy to the HF result. It will be interesting to compute nuclei on lattices with finite-range interactions, as done in the quantum Monte Carlo simulations of nuclear lattice effective field theory~\cite{elhatisari2016,elhatisari2024}. Those works treat an SU(4) symmetric interaction non-perturbatively and add all remaining interaction contributions in first-order perturbation theory. Clearly, for zero-range interactions the present results show that the mean field (or coupled cluster with singles) gives most of the energy.

For completeness we return to the lattice reference states and compare the ground-state energies of $^{3,4}$He computed with full configuration interaction, IMSRG, and CCSD for $L=3$ in Table~\ref{tab:IMSRGresults}. 
For $^4$He the IMSRG(2) and CCSD results both miss up to a few MeV in energy, and the difference to full configuration interaction increases with decreasing lattice spacing. We interpret this as follows. The finite-size corrections increase with decreasing lattice spacing and the corresponding correlations, i.e., tunneling from one box to another,  involve the whole nucleus (or its center-of-mass coordinate). Finite size corrections, i.e., three-particle--three-hole correlations are particularly large for $^3$He because that nucleus is more weakly bound than the $\alpha$ particle. Thus IMSRG(2) and CCSD miss quite a bit of the $^3$He ground-state energy for $a=2$ and 1.7~fm.
It is also worth noting that the cutoffs of our Hamiltonians increase with decreasing lattice spacing, with $\Lambda = \pi / a$.
The Hamiltonians with $a=2.0$ and especially 1.7~fm are clearly harder, which makes the many-body calculations more challenging.
We also note that for the latter case the lattice reference state is unbound.

\begin{table}
\setlength{\tabcolsep}{3pt}
\centering
\caption{Ground-state energies (in MeV) of the $^{3,4}$He nuclei for a lattice with extent $L=3$ and spacing $a$ (in fm), computed with FCI, the IMSRG(2), and CCSD.}
\label{tab:IMSRGresults}
\begin{tabular}{crrrrrr}
\hline\noalign{\smallskip}
\multirow{2}{*}{$a$} & \multicolumn{3}{c}{$^3$He} & \multicolumn{3}{c}{$^4$He} \\
& \multicolumn{1}{c}{FCI}   & \multicolumn{1}{c}{IMSRG(2)}  & \multicolumn{1}{c}{CCSD}   & \multicolumn{1}{c}{FCI}   & \multicolumn{1}{c}{IMSRG(2)}   & \multicolumn{1}{c}{CCSD}\\ 
\noalign{\smallskip}\hline\noalign{\smallskip}
2.5 & $-17.40$ & $-15.43$ & $-15.70$ & $-31.10$ & $-29.80$ & $-30.09$\\
2.0 & $-17.65$ & $-12.51$ & $-13.23$ & $-32.99$ & $-29.15$ & $-29.83$\\
1.7 & $-16.26$ &  $-5.28$ &  $-6.68$ & $-34.49$ & $-28.45$ & $-29.37$\\
\hline\noalign{\smallskip}
\end{tabular}
\end{table}

\subsection{Results for $^8$Be, $^{12}$C, and $^{16}$O}
\label{sec:nuclei}

\begin{figure}[htb]
	\centering
	\includegraphics[width=1.0\columnwidth]{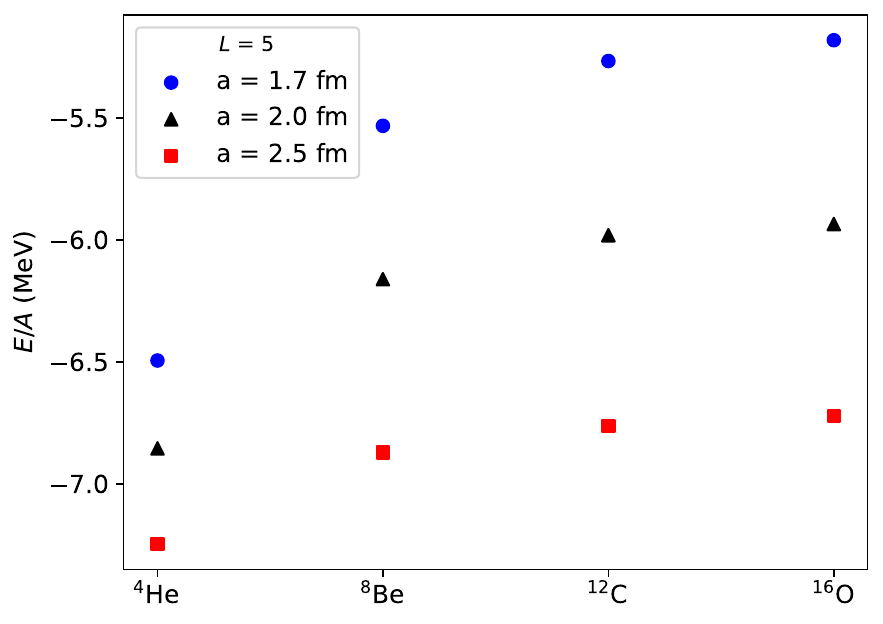}
	\caption{Ground state energy per nucleon for $^4$He, $^8$Be, $^{12}$C, and $^{16}$O for three difference lattice spacings done with a lattice size $L=5$, computed with coupled cluster with singles and  doubles.}\label{fig:a_latE}
\end{figure}

In this section we compute the ground-state energies of $^8$Be, $^{12}$C, and $^{16}$O. 
Figure~\ref{fig:a_latE} shows the ground-state energies per nucleon, computed with coupled cluster with singles and doubles using the different interactions of Table~\ref{tab:results} on lattices with $L=5$. 

Clearly, the nuclei $^8$Be, $^{12}$C, and $^{16}$O, which consist of multiple $\alpha$ particles, are not bound with respect to $\alpha$ decay. This short-coming of pion-less effective field theory is known also from  other studies~\cite{contessi2017,bansal2018,bazak2019,lu2019} and particularly simple to understand for our lattice implementation of this effective theory. As the interactions are onsite only, i.e., they have zero range, the potential cannot bind nuclei beyond the $\alpha$ particle. For the reference states of the nuclei $^8$Be, $^{12}$C, and $^{16}$O we chose $\alpha$ particles on neighboring lattice sites in the configurations determined in Refs.~\cite{epelbaum2011,epelbaum2012,epelbaum2014}. We confirmed that the energies of these nuclear systems become very close to those of multiple $\alpha$ particles for reference states that consist of $\alpha$ particles on well separated lattice sites.     

\begin{figure}[htb]
	\centering
	\includegraphics[width=1.0\columnwidth]{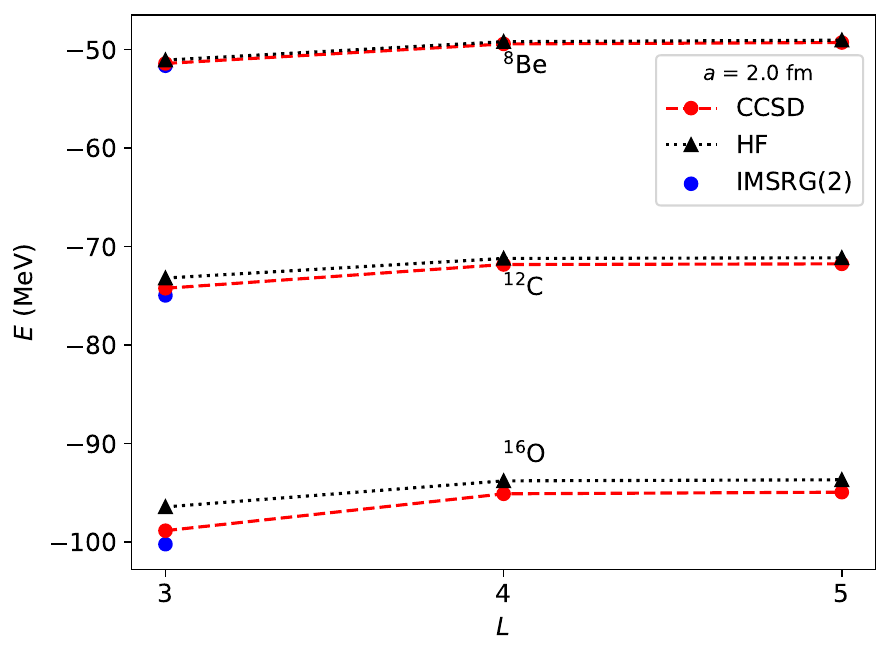}
	\caption{Ground-state energies of $^8$Be, $^{12}$C, and $^{16}$O as a function of lattice size $L$ for the spacing $a=2$~fm, computed with Hartree Fock (HF), coupled cluster with singles and doubles (CCSD) based on a lattice reference state, and IMSRG(2) based on a Hartree-Fock reference state.}\label{fig:LSize}
\end{figure}

Figure~\ref{fig:LSize} shows that the energy converges rapidly as the size of the lattice is increased. For $^{16}$O, for instance, the CCSD ground-state energy changes by less than 0.2~MeV from $L=4$ to 5. We see again that the Hartree-Fock energies are close to those from coupled cluster with singles and doubles. For illustration, we also show IMSRG(2) calculations for $^{8}$Be, $^{12}$C, and $^{16}$O based on a Hartree-Fock reference state for $L=3$. The results are consistent with the CCSD results starting from a lattice reference state, which reflects the fact that HF and CCS are the dominant contributions to the binding energy. The IMSRG(2) gives only small contributions to the binding energy beyond Hartree Fock.

We saw that the patterns observed in the $A=2,3$, and 4 nucleon systems are also reflected in other light nuclei such as $^8$Be, $^{12}$C, and $^{16}$O that consists of multiple $\alpha$ particles. Hartree Fock (or coupled-cluster singles) yield most of the energy on the lattice and lattices of size $L=4$ or 5 (with spacings around $a\approx 2$~fm) are large enough for sufficiently well converged results. In contrast, nuclei such as $^6$He (which can be thought of as the bound state of an $\alpha$ particle and a dineutron) or $^6$Li (which can be thought of as a deuteron-$\alpha$ bound state) cannot be accurately computed with coupled-cluster with singles and doubles. We think that this is a peculiarity of the zero-range interactions we used in this work.

\section{Outlook}
\label{sec:outlook}

This package evolved from a research project and has also been used in education. The recent paper~\cite{gu2025} on quantum computing of atomic nuclei used routines from \texttt{NuLattice} to generate lattice Hamiltonians from pion-less effective field theory and to benchmark with results from full configuration interaction. Recent coupled-cluster computations used routines from \texttt{NuLattice} and reported on the exactness of the normal-ordered two-body approximation for zero-range three nucleon forces~\cite{rothman2025}. In the 2025 summer school ``Emergence of Collective Motion in Atomic Nuclei'' of the FRIB Theory Alliance, \texttt{NuLattice} routines from full configuration interaction were used by students to learn about (i) the convergence patterns of light nuclei with respect to increasing lattice sizes, (ii) the renormalization and ``running'' of coupling constants in pion-less effective field theory through variation of the lattice spacing, and (iii) the need of a three-body contact~\cite{bedaque1999}. While \texttt{NuLattice} now contains several person-months of work and about 4500 lines (including doc strings) of Python code, the development and implementation of its routines was much simpler and faster than for \textit{ab initio} computations that use the harmonic oscillator basis. Most importantly, interesting and useful computations can be performed on laptops.  

Much work remains. Finite-range interactions~\cite{elhatisari2016,lu2019} and interactions from chiral effective field theory~\cite{elhatisari2024} are lacking; higher precision in coupled-cluster and IMSRG computations is desirable; a sparse solution of the IMSRG equations would enable calculations on large lattice sizes; observables other than the ground-state energy are of interest; excited states need to be computed. The list goes on and on. 

We invite readers to join us in this effort! \texttt{NuLattice} is documented, and the authors of each module have been identified and credited. This makes it easy to contribute to \texttt{NuLattice} and to receive recognition for one's work.

\section{Summary}
\label{sec:summary}
This paper introduced the software package \texttt{NuLattice}, a Python package of computational tools for \textit{ab initio} computations of nuclei on lattices. Using \texttt{NuLattice}, we performed \textit{ab initio} computations of nuclei on lattices, using Hartree Fock, coupled-cluster theory, the in-medium similarity renormalization group, and full configuration interaction. The employed Hamiltonians were from pion-less effective field theory at leading-order and consisted of two-body and three-body contact interactions. We found that Hartree Fock and coupled-cluster with singles yielded the bulk of the ground-state energy for compact nuclei. We confirmed that pion-less effective field theory fails to bind $\alpha$ particles into nuclei. We hope that \texttt{NuLattice} might be useful to researchers and educators and invite readers to contribute. 

\begin{acknowledgements}
We thank Francesca Bonaiti, Tor Dj\"arv, Bingcheng He, and Gustav Jansen for many stimulating discussions and Serdar Elhatisari and Dean Lee for useful communications.
This work was supported by the U.S.\ Department of Energy, Office of
Science, Office of Nuclear Physics, under Award Nos.~DE-FG02-96ER40963; by the U.S.\ Department of Energy, Office of Science, Office of Advanced Scientific Computing Research and Office of Nuclear Physics, Scientific Discovery through Advanced Computing (SciDAC) program (SciDAC-5 NUCLEI), and by the Laboratory Directed Research and Development Program of Oak Ridge National Laboratory (ORNL). 
This research used resources of the Oak Ridge Leadership Computing Facility at the Oak Ridge National Laboratory, which is supported by the Office of Science of the U.S.\ Department of Energy under Contract No.~DE-AC05-00OR22725.
\end{acknowledgements}

\bibliographystyle{apsrev4-1-mh-mod}
\bibliography{master}

\end{document}